\documentclass[preprint,12pt]{elsarticle}
\journal{Astroparticle Physics}

\usepackage{lineno} 
\usepackage{graphicx}
\usepackage{amsmath,amssymb,bbold,bm}
\usepackage{epstopdf}
\usepackage[T1]{fontenc}

\begin{document}
\def\nuc#1#2{${}^{#1}$#2}
\def\mee{$\langle m_{\beta\beta} \rangle$}
\def\mnu{$m_{\nu}$}
\def\ml{$m_{lightest}$}
\def\gnu{$\langle g_{\nu,\chi}\rangle$}
\def\mmod{$\| \langle m_{\beta\beta} \rangle \|$}
\def\mb{$\langle m_{\beta} \rangle$}
\def\BBz{$\beta\beta(0\nu)$}
\def\BBm{$\beta\beta(0\nu,\chi)$}
\def\BBt{$\beta\beta(2\nu)$}
\def\nonubb{$\beta\beta(0\nu)$}
\def\twonubb{$\beta\beta(2\nu)$}
\def\BB{$\beta\beta$}
\def\Mz{$M_{0\nu}$}
\def\Mt{$M_{2\nu}$}
\def\MzG{$M^{GT}_{0\nu}$}           %Gamov-Teller
\def\MzF{$M^{F}_{0\nu}$}                %Fermi
\def\MtG{$M^{GT}_{2\nu}$}           %Gamov-Teller
\def\MtF{$M^{F}_{2\nu}$}                %Fermi
\def\Gz{$G_{0\nu}$}					%phase space factor for 0nu
\def\Tz{$T^{0\nu}_{1/2}$}
\def\Tt{$T^{2\nu}_{1/2}$}
\def\Tc{$T^{0\nu\,\chi}_{1/2}$}
\def\Rz{$\Gamma_{0\nu}$}            %0 nu decay rate
\def\Rt{$\Gamma_{2\nu}$}            %2 nu decay rateq
\def\ms{$\delta m_{\rm sol}^{2}$}
\def\ma{$\delta m_{\rm atm}^{2}$}
\def\mot{$\delta m_{12}^{2}$}
\def\mtt{$\delta m_{23}^{2}$}
\def\ts{$\theta_{\rm sol}$}
\def\ta{$\theta_{\rm atm}$}
\def\ttwo{$\theta_{12}$}
\def\tot{$\theta_{13}$}
\def\gpp{$g_{pp}$}                  % the g_pp of QRPA fame
\def\gA{$g_{A}$}                  % the Axial Vector coupling constant
\def\qval{$Q_{\beta\beta}$}                 % The Q-value
\def\be{\begin{equation}}
\def\ee{\end{equation}}
\def\cpRty{cnts/(ROI-t-y)}
\def\onecpRty{1~count/(ROI-t-y)}
\def\threecpRty{3~counts/(ROI-t-y)}
\def\ppc{P-PC}                          % P-type Point Contact
\def\nsc{N-SC}                          % N-type Segmented Contact
\def\cosixty{$^{60}Co$}
\def\thttt{$^{232}\mathrm{Th}$}
\def\utte{$^{238}\mathrm{U}$}
\def\mubqkg{$\mu\mathrm{Bq/kg}$}
\def\cusulfate{$\mathrm{CuSO}_4$}
\def\MJ{{\sc Majorana}}             %Majorana project name
\def\DEM{{\sc Demonstrator}}             %Demonstrator in small caps
\def\MJDEMbf{\bfseries{\scshape{Majorana Demonstrator}}}
\def\MJbf{\bfseries{\scshape{Majorana}}}
\def\MJDEMit{\itshape{\scshape{Majorana Demonstrator}}}
\newcommand{\Gerda}{GERDA}
\newcommand{\GT}{GEANT-3}
\newcommand{\GF}{Geant4}
\newcommand{\MaGe}{\textsc{MaGe}}

\begin{frontmatter}

\title{Muon Flux Measurements at the Davis Campus of the Sanford Underground Research Facility with the {\sc Majorana Demonstrator} Veto System}

%{\blhill} %OK 9/29/14
%{\duke} 	 	 %OK 9/12/14
%{\ITEP}  	%OK 9/25/14
%{\JINR} 
%{\lanl} 	 %OK 9/12/14
%{\lbnl}  	%OK 9/25/14	
%{\ornl}  %OK 9/12/14 	
%{\ou} %OK 9/22/14
%{\pnnl}  %OK 9/12/14
%{\sdsmt}  %OK 9/12/14
%{\tunl} x
%{\ttu}  	%OK 9/25/14
%{\alberta}  	%OK 9/25/14
%%{\ucne}  x
%%{\ucph}  x
%%{\uchic} x
%{\unc} 	 	%OK 9/25/14
%{\usc} 	%OK 9/25/14
%{\usd} 	%OK 9/16/14
%{\ut} %OK 9/12/14
%{\uw} 	%OK 9/16/14

\author[lbnl]{N.~Abgrall}
\author[pnnl]{E.~Aguayo}
\author[usc,ornl]{F.T.~Avignone~III}
\author[ITEP]{A.S.~Barabash}
\author[ornl]{F.E.~Bertrand}
\author[lbnl]{A.W.~Bradley}
\author[JINR]{V.~Brudanin}
\author[duke,tunl]{M.~Busch}
\author[uw]{M.~Buuck}
\author[usd]{D.~Byram}
\author[sdsmt]{A.S.~Caldwell}
\author[lbnl]{Y-D.~Chan}
\author[sdsmt]{C.D.~Christofferson}
\author[lanl]{P.-H.~Chu}
\author[uw]{C. Cuesta}
\author[uw]{J.A.~Detwiler}
\author[sdsmt]{C. Dunagan}
\author[ut]{Yu.~Efremenko \corref{cor1}}
\author[ou]{H.~Ejiri}
\author[lanl]{S.R.~Elliott}
\author[ornl]{A.~Galindo-Uribarri}
\author[unc,tunl]{T.~Gilliss}
\author[unc,tunl]{G.K.~Giovanetti}
\author[lanl]{J. Goett}
\author[ncsu,ornl,tunl]{M.P.~Green}
\author[uw]{J. Gruszko}
\author[uw]{I.S.~Guinn}
\author[usc]{V.E.~Guiseppe}
\author[unc,tunl]{R.~Henning}
\author[pnnl]{E.W.~Hoppe}
\author[sdsmt]{S. Howard}
\author[unc,tunl]{M.A.~Howe}
\author[usd]{B.R.~Jasinski}
\author[blhill]{K.J.~Keeter}
\author[ttu]{M.F.~Kidd}
\author[ITEP]{S.I.~Konovalov}
\author[pnnl]{R.T.~Kouzes}
\author[pnnl]{B.D.~LaFerriere}
\author[uw]{J. Leon}
\author[ut]{A.M.~Lopez}
\author[unc,tunl]{J.~MacMullin}
\author[usd]{R.D.~Martin\fnref{MartinsPerm}}
\author[lanl]{R. Massarczyk}
\author[unc,tunl]{S.J.~Meijer}
\author[lbnl]{S.~Mertens}
\author[pnnl]{J.L.~Orrell}
\author[unc,tunl]{C. O'Shaughnessy}
\author[pnnl]{N.R.~Overman}
\author[lbnl]{A.W.P.~Poon}
\author[ornl]{D.C.~Radford}
\author[unc,tunl]{J.~Rager}
\author[lanl]{K.~Rielage}
\author[uw]{R.G.H.~Robertson}
\author[ut,ornl]{E. Romero-Romero}
\author[lanl]{M.C.~Ronquest}
\author[lbnl]{C.~Schmitt}
\author[unc,tunl]{B. Shanks}
\author[JINR]{M.~Shirchenko}
\author[usd]{N.~Snyder}
\author[sdsmt]{A.M.~Suriano}
\author[usc]{D.~Tedeschi}
\author[unc,tunl]{J.E.~Trimble}
\author[ornl]{R.L.~Varner}
\author[JINR]{S.~Vasilyev}
\author[lbnl]{K.~Vetter\fnref{lbnleng}}
\author[unc,tunl]{K.~Vorren}
\author[ornl]{B.R.~White}
\author[unc,tunl,ornl]{J.F.~Wilkerson}
\author[usc]{C. Wiseman}
\author[lanl]{W.~Xu\fnref{XuPerm}}
\author[JINR]{E.~Yakushev}
\author[ornl]{C.-H.~Yu}
\author[ITEP]{V.~Yumatov}
\author[JINR]{I.~Zhitnikov}
			
\author{\newline (The {\sc Majorana} Collaboration)}
\cortext[cor1]{Corresponding author, yefremen@utk.edu, phone: +1 865 974 7857, fax: +1 865 974 7843}

%These have to be in order in which they appear for the authors:
\address[lbnl]{Nuclear Science Division, Lawrence Berkeley National Laboratory, Berkeley, CA, USA}
\address[pnnl]{Pacific Northwest National Laboratory, Richland, WA, USA}
\address[usc]{Department of Physics and Astronomy, University of South Carolina, Columbia, SC, USA}
\address[ornl]{Oak Ridge National Laboratory, Oak Ridge, TN, USA}
\address[ITEP]{National Research Center ``Kurchatov Institute'' Institute for Theoretical and Experimental Physics, Moscow, Russia}
\address[JINR]{Joint Institute for Nuclear Research, Dubna, Russia}
\address[duke]{Department of Physics, Duke University, Durham, NC, USA}
\address[tunl]{Triangle Universities Nuclear Laboratory, Durham, NC, USA}
\address[uw]{Center for Experimental Nuclear Physics and Astrophysics, and Department of Physics, University of Washington, Seattle, WA, USA}
\address[usd]{Department of Physics, University of South Dakota, Vermillion, SD, USA} 
\address[sdsmt]{South Dakota School of Mines and Technology, Rapid City, SD, USA}
\address[lanl]{Los Alamos National Laboratory, Los Alamos, NM, USA}
\address[ut]{Department of Physics and Astronomy, University of Tennessee, Knoxville, TN, USA}
\address[ou]{Research Center for Nuclear Physics and Department of Physics, Osaka University, Ibaraki, Osaka, Japan}
\address[unc]{Department of Physics and Astronomy, University of North Carolina, Chapel Hill, NC, USA}
\address[ncsu]{Department of Physics, North Carolina State University, Raleigh, NC, USA}
\address[blhill]{Department of Physics, Black Hills State University, Spearfish, SD, USA} 
\address[ttu]{Tennessee Tech University, Cookeville, TN, USA}
\fntext[lbnleng]{Alternate Address: Department of Nuclear Engineering, University of California, Berkeley, CA 94720}
\fntext[XuPerm]{Present Address: Department of Physics, University of South Dakota, Vermillion, SD, USA}
\fntext[MartinsPerm]{Present Address: Department of Physics, Engineering Physics and Astronomy, Queen's University, Kingston, Canada}

\begin{abstract}
We report the first measurement of the total muon flux underground at the Davis Campus of the Sanford Underground Research Facility at the 4850 ft level. Measurements were performed using the \MJ\ \DEM\ muon veto system arranged in two different configurations. The measured total flux is $(5.31\pm0.17)\times10^{-9}$ $\mu$/s/cm$^2$. 

\end{abstract}

\end{frontmatter}

\section{Introduction}
The Davis Campus at the Sanford Underground Research Facility \\(SURF)~\cite{Heise2014}, located in the former Homestake gold mine, is situated at a depth of 4850 ft near the city of Lead, SD, USA. SURF has become a prime site for  low background science in the United States since the inauguration of its Davis Campus in 2012. Accurate characterization of the muon flux and average rock density is important for understanding cosmic-ray-induced backgrounds not only in existing experiments presently deployed at SURF, but also for future projects. A previous measurement of the vertical muon flux at the 4850-ft level has been reported~\cite{Cherry1983}, and the total muon flux was measured for the 800- and 2000-ft levels~\cite{Gray2011} at SURF. The total muon flux at the 4850-ft level was calculated to be $(4.4 \pm 0.1)\times10^{-9}$ $\mu$/s/cm$^2$~\cite{Mei2006}. In this article, we present a first measurement of the total muon flux at the 4850-ft level using the \MJ\ \DEM\ muon veto system. We compare our measurement to previous work, and to our own simulation of muon transport from the surface to the experiment using geological measurements of the average rock density of the SURF overburden.

The \MJ\ \DEM\ is an array of enriched and natural high-purity germanium (HPGe) detectors that are used to search for the zero neutrino double-beta (\BBz) decay of the isotope \nuc{76}{Ge}. The details of the experiment's design are given in Ref.~\cite{Abgrall2014} and only key aspects required for this result are discussed here. The specific goals of the \MJ\ \DEM\ are to:

\begin{enumerate}
\item Demonstrate a path forward to achieving a background rate at or below \onecpRty\ in the 4-keV region of interest (ROI) around the  2039-keV  Q-value  for \nuc{76}{Ge} \BBz\ decay. This is required for tonne-scale germanium-based searches that will probe the inverted-ordering neutrino-mass parameter space for the effective Majorana neutrino mass in \BBz\ decay.
\item Show technical and engineering scalability toward a tonne-scale instrument.
\item Perform searches for additional physics beyond the Standard Model, such as dark matter and axions.
\end{enumerate}

The \MJ\ Collaboration has designed a modular instrument composed of two cryostats built from ultra-pure electroformed copper, with each cryostat capable of housing over 20 kg of HPGe detectors. The \MJ\ \DEM\  contains 30 kg of detectors fabricated from Ge material enriched to 88\% in \nuc{76}{Ge} and another 15 kg fabricated from natural Ge (7.8\% \nuc{76}{Ge}). The modular approach allows us to assemble and optimize each cryostat independently, providing a fast deployment with minimal effect on already-operational detectors. 

\begin{figure}[ht]
\begin{center}
\includegraphics[width=8cm]{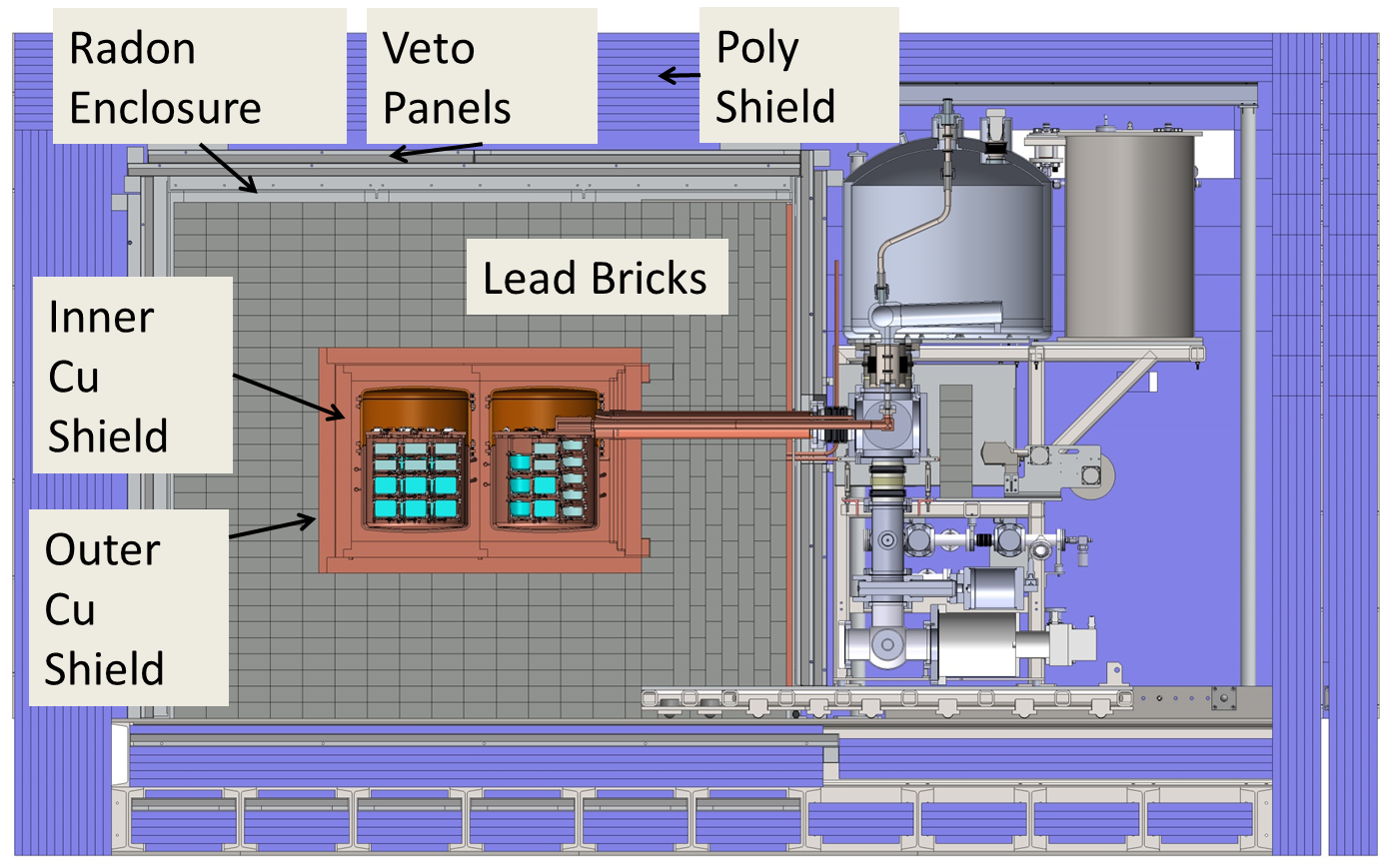}
\caption{Color online: The shield system in cross section, shown with both cryostats installed.}
\label{fig:ShieldOverview}
\end{center}
\end{figure}

Starting from the innermost cavity, the cryostats are surrounded by a compact graded shield composed of an inner layer of electroformed copper, a layer of commercially sourced C10100 copper, high-purity lead, an active muon veto, borated polyethylene, and pure polyethylene shielding. The cryostats, copper, and lead shielding are enclosed in a radon exclusion box and rest on an over-floor table that has openings for the active muon veto and polyethylene shielding panels situated below the detector. The entire experiment is located in a clean room at the 4850 ft level of SURF. A high-level summary of shield components is shown in Fig.~\ref{fig:ShieldOverview}. 

A large fraction of the plastic scintillator panels comprising the active muon-veto system were operated in different configurations at the experimental site during Ge detector constructions and commissioning. We used the resulting data to measure the total muon flux at the Davis Campus at SURF for the first time.

%\section{The \MJDEMbf\ Veto System}
\section{The \MJDEMbf\ Muon Veto System}
\begin{figure}[ht]
\begin{centering}
\begin{minipage}[b]{0.4\textwidth}
\includegraphics[width=\textwidth]{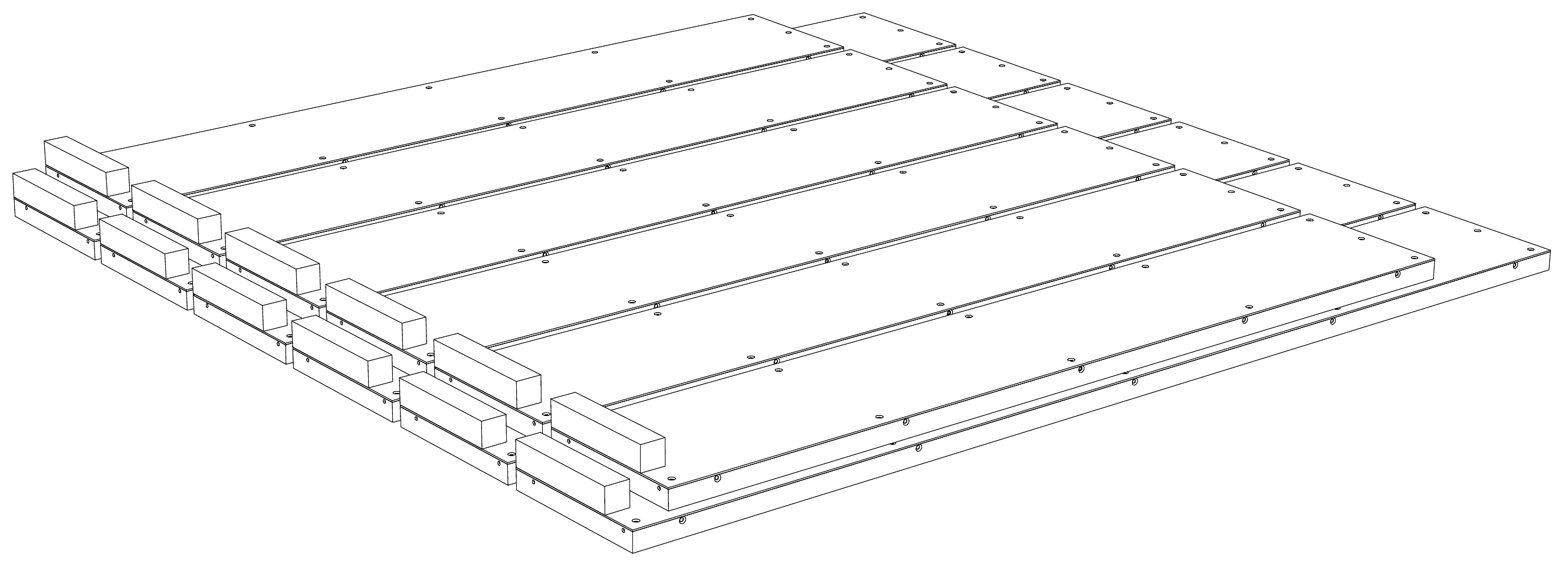}
\end{minipage}
\begin{minipage}[b]{0.4\textwidth}
\includegraphics[width=\textwidth]{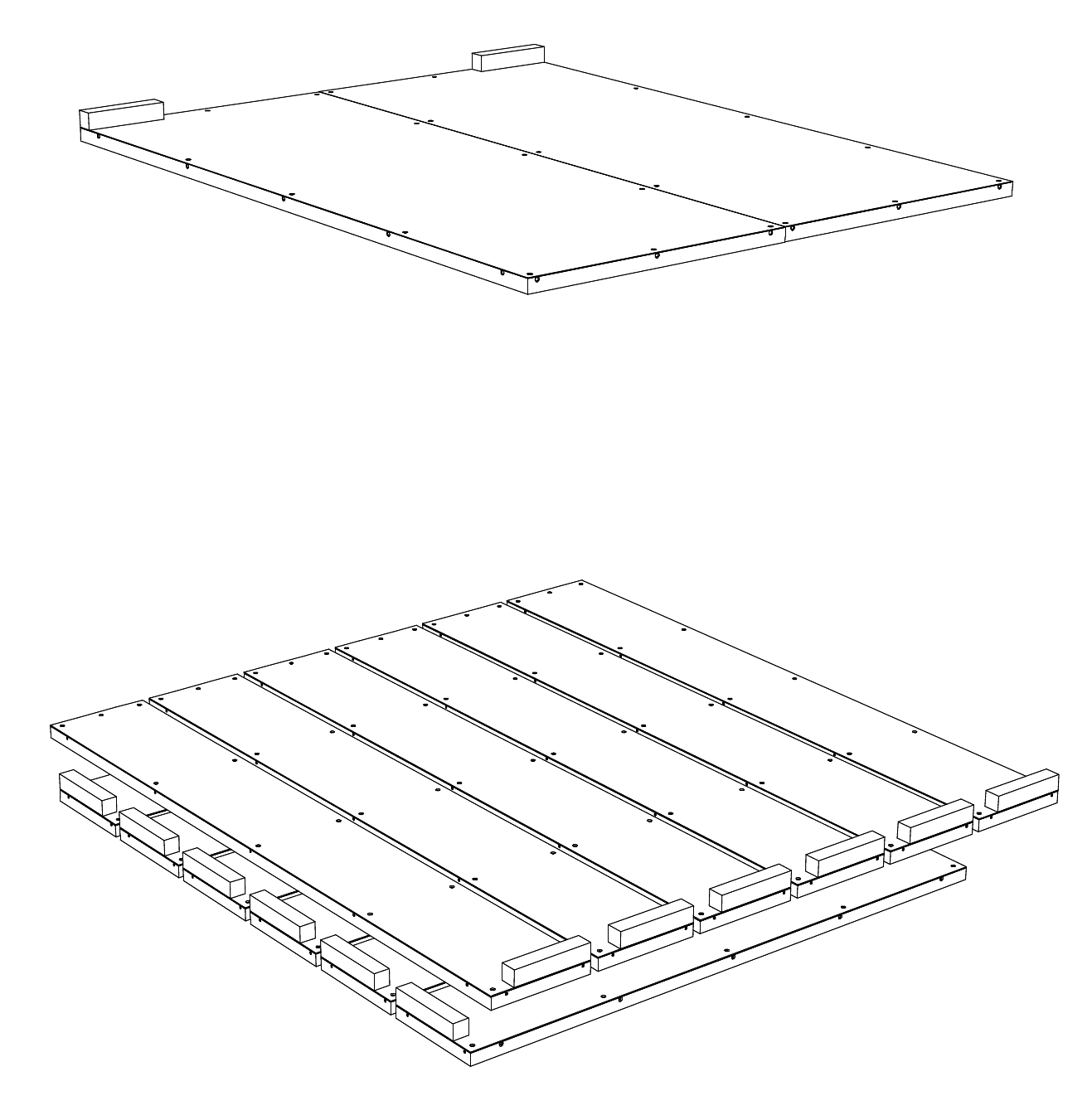}
\end{minipage}
\caption{Layouts of the {\sc Majorana Demonstrator} veto panels used in this study. The layout on the left shows the two-fold coincidence arrangement, and on the right is the three-fold coincidence arrangement. Muon selection require a hit in at least one panel in each layer. For the three-fold coincidence arrangement massive lead shielding was present between the top layer and the upper of the bottom two layers as shown in Fig.~\ref{fig:ShieldOverview}. The bottom layers reside within a steel support over-floor table, which is not shown. All other {\sc Demonstrator} components are also suppressed in this view.}
\label{fig:VetoOverview}
\end{centering}
\end{figure}

The  \MJ\ \DEM\ muon veto system was designed to completely enclose the passive copper and lead shield within two layers of scintillating panels while minimizing gaps. Each layer is composed of 2.54-cm-thick EJ-204B scintillating acrylic sheets encapsulated within Al cladding. These detector panels have various shapes and dimensions resulting in a total area of $\sim$37~m$^2$. The \DEM\ uses a total of 32 veto panels, including twelve that reside within openings of the overfloor table in two orthogonal orientations. The data presented in this paper is based on the operation of two configurations, one with 12 veto panels requiring two-fold coincidence, and one with 14 veto panels requiring three-fold coincidence.  The arrangement of the veto panels used for each configuration is shown in Fig.~\ref{fig:VetoOverview}. More details on each configuration are given in Sections 3 and 4 below.

Light from each individual panel was read out by a single 1.27-cm photomultiplier tube (PMT) with wavelength shifting fibers embedded into grooves machined in the scintillator. The panel components were optimized to provide high light output, good light collection uniformity, and excellent muon-detection efficiency ($\epsilon_D>$99.9\%)~\cite{Bugg2014}. The details of the data acquisition system for the veto system were given in Ref.~\cite{Abgrall2014}. Performance of each panel is constantly monitored with Light Emitting Diodes (LED) embedded in the scintillator. Reconstructed LED events were also used to measure the live time of the system. The LEDs are pulsed at a frequency known with a precision of 0.1\%.

In a deep underground laboratory the muon flux is low, but $\gamma$ rays from the experimental apparatus and the laboratory environment are significant. As escribed in \cite{Bugg2014}, the use of the relatively thin 2.54-cm scintillator panels presents certain challenges for separating muons from $\gamma$ rays and random $\gamma$-ray coincidences at the SURF depth. The most probable muon-energy deposition in the veto panels is $\sim$5~MeV, which is low enough that the high energy tail of the $\gamma$-ray energy distribution can potentially encroach upon the muon peak, potentially overwhelming the muon contribution to the spectrum. 
The design and construction of the veto panels achieved good light collection and ensured that the $\mu$ peak remained well-separated from the $\gamma$-ray tail for two-fold or higher-multiplicity
coincidences, even at the low muon flux of the Davis Campus.

\section{Two-Fold Coincidence Measurement}
\begin{figure}[ht]
\begin{center}
\includegraphics[width=9cm]{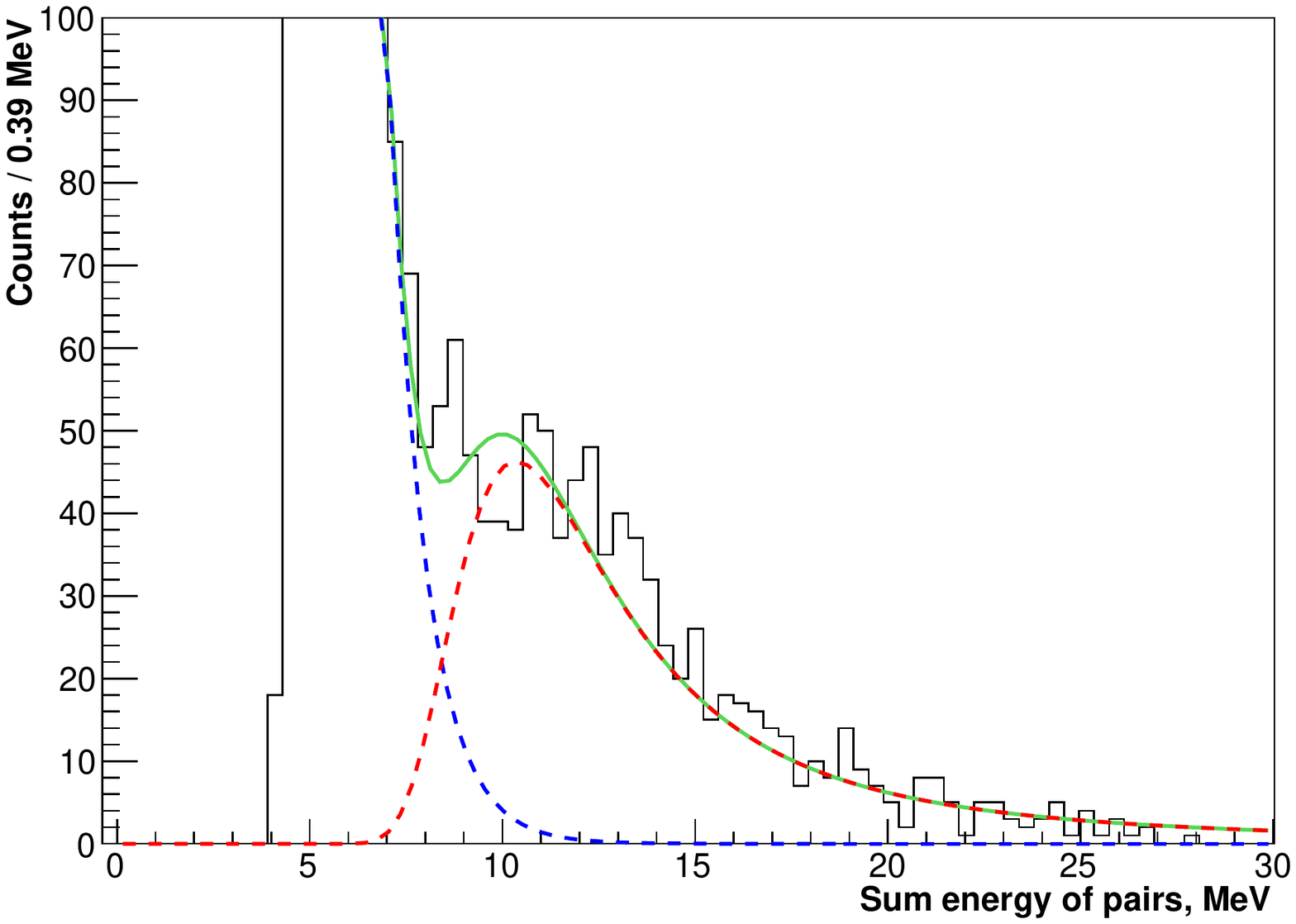}
\caption{Color online: The energy deposition of selected events for the two-fold coincidence configuration (black solid histogram).  The horizontal scale is the summed energy deposition of the paired panels. The tail from the energy deposition of the $\gamma$ rays (blue dashed curve) is fitted with an exponential distribution. The signal from muons is fitted by a Landau distribution (red dashed curve). The total fit is given by the solid green curve. The most probable summed energy deposit value is 10.7 $\pm$ 0.2 MeV.}
\label{fig:EnergyDeposition}
\end{center}
\end{figure}

For the first configuration we used the twelve narrow bottom panels arranged in six pairs. Prior to installation into their final location, six panels, each with dimensions of $32\times182$ cm$^2$, were placed parallel to and on top of an additional six panels with dimensions $32\times223$ cm$^2$. We selected events where both a top and bottom panel simultaneously generated a signal above 1.8 MeV. In this two-fold coincidence configuration, the live time is 1536 h ($5.53\times10^6$ s) between December 19, 2013 and March 11, 2014. The sum of energy deposits in the two panels is shown in Fig.~\ref{fig:EnergyDeposition}. From the figure, one can see that the tail from the $\gamma$ rays makes it difficult to precisely measure the muon flux from this configuration. Data were fit by combination of an exponential tail approximating the  $\gamma$ background (blue line), and a Landau distribution for muons (red line). The characterization of the $\gamma$ background tail with an exponential function is justified through an independent fit to accidental two-fold coincidences between the bottom panels. The extracted number of muons passing through system is 912 $\pm$ 43. We note that because the pairs of panels were adjacent, this configuration is sensitive to the total muon flux but not the muon angular distribution.

The individual data runs were 8 hours and the spread in the number of detected events per run follows Poisson statistics. All six detector pairs have similar muon rates that agree within statistical fluctuations.

\section{Three-Fold Coincidence Measurement }
For the second configuration we used the veto panels placed in their planned final arrangement. In this configuration, data were selected for three-fold coincidences. Two of these signals came from each of the two layers of twelve panels (arranged in their final six by six orthogonal configuration in the over-floor, as indicated in Fig.~\ref{fig:VetoOverview}), and the third signal came from one of two large panels mounted on the top of the experiment's passive shielding. A 1.6-m tall lead shield is situated between the top and bottom panels with a small central cavity of dimensions of ($90 \times 50 \times 60$ cm$^3$). The top panels are located side by side and their dimensions are each $84\times 211$ cm$^2$. 

\begin{figure}[ht]
\begin{center}\includegraphics[width=9cm]{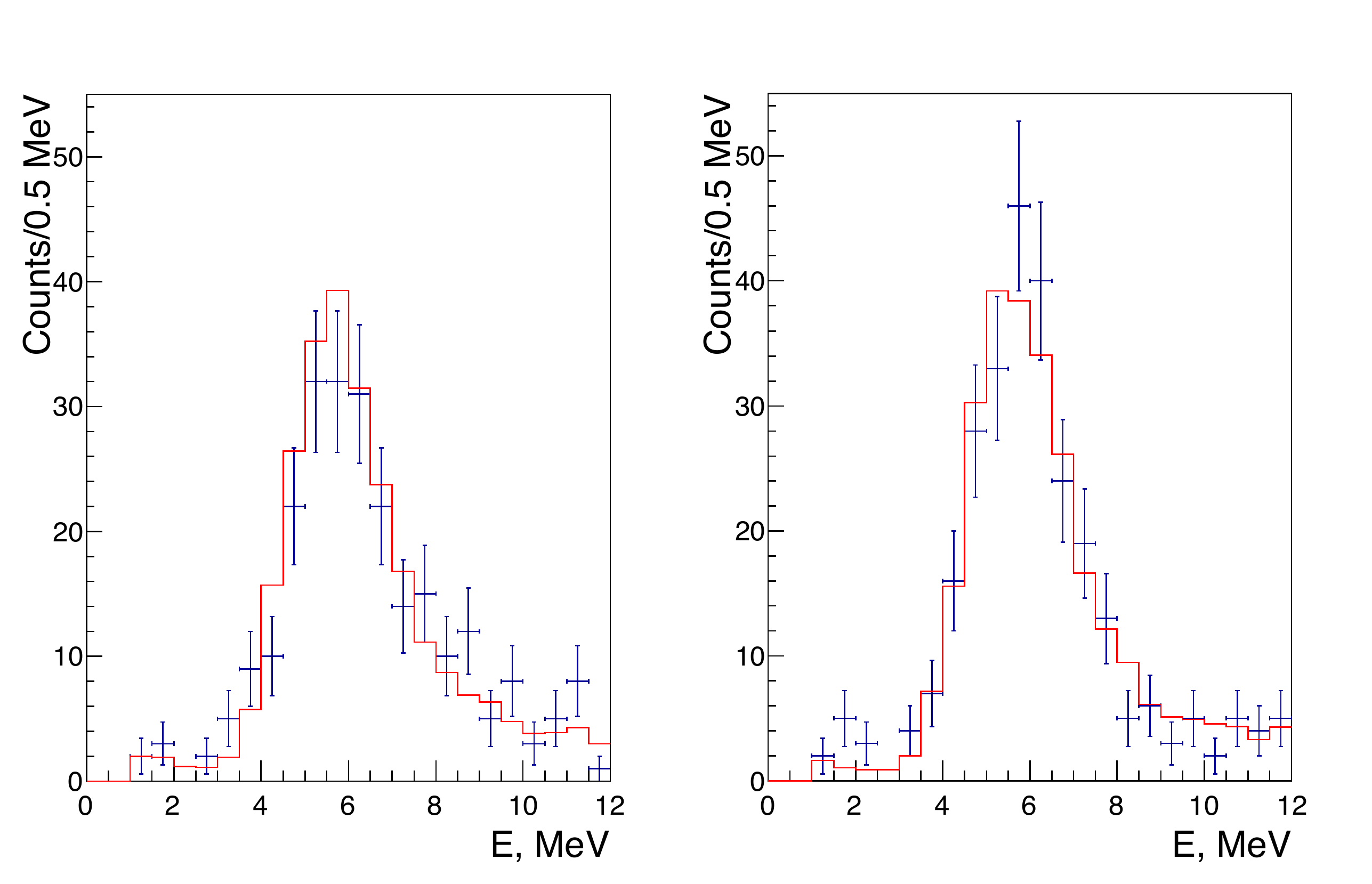}
\caption{Color online: Energy distributions in the two large top large panels for the second configuration during a three-fold coincidence, showing a clear muon signal. The left figure shows data from the upper left panel in coincidence with bottom panels and on the right is data from the upper right panel in coincidence with bottom panels. The solid line is simulation, which has much larger statistics.}
\label{fig:TopTwo}
\end{center}
\end{figure}

In this configuration, the live time was 2678 h ($9.64\times10^6$ s) collected between June 20 and November 10, 2014, during which a total of 615 $\pm$ 25 muons were detected. For this triple-plane configuration, the random $\gamma$-ray background is negligible and a Landau distribution of muon energy deposition in the panel can be clearly seen in Fig.~\ref{fig:TopTwo}. Based on these data we were able to verify the energy calibration of all panels by reconstruction of the muon peak. The shape of the Landau distribution was consistent between the two- and three-fold configurations within experimental uncertainties. The run-to-run event variations agreed within Poisson fluctuations. 

\section{Muon Simulations and Results}
To estimate the total muon flux we must estimate the effective cross-sectional area of our detector configurations relative to the muon angular distribution at the Davis Campus. Since the two configurations have qualitatively different response to the muon angular distribution, the difference between the extracted flux values provides a cross-check on the sensitivity to the details of the assumed angular distribution. To model the muon angular distribution and the response of each configuration to it, we simulated muons propagating from the surface through the rock to the Davis Campus, and then through the
 \MJ\ \DEM\  laboratory and the detectors. 
  
\begin{figure}[ht]
\begin{center}
\includegraphics[width=0.85\textwidth]{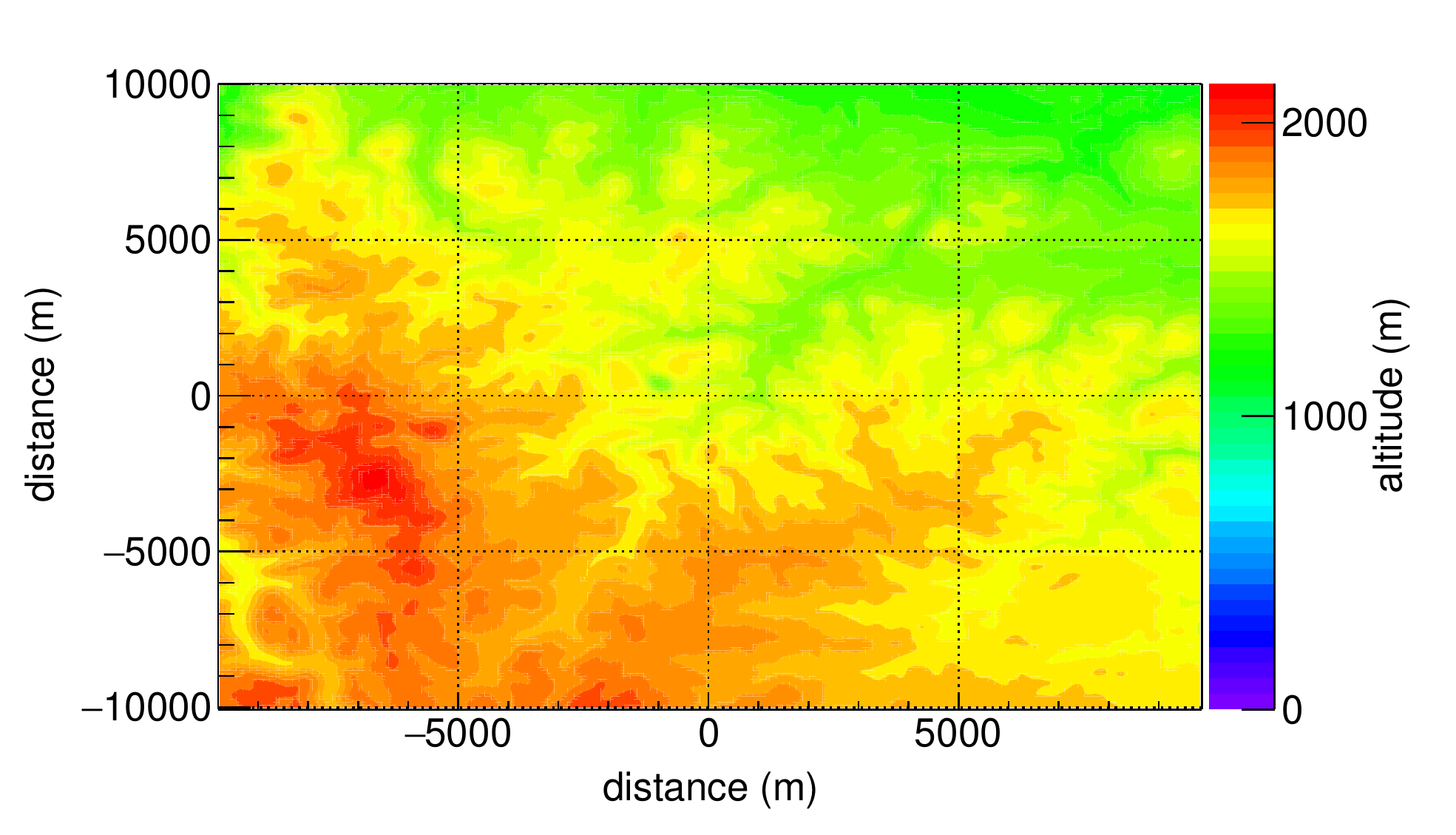}
\caption{The altitude in m of the surface directly above the underground laboratory, which is located at the origin of the plot at an altitude of 119 m. Geographic North is directed towards the top of the figure. Source for this plot is from\cite{shuttle} 
}
\label{fig:landscape}
\end{center}
\end{figure}

To understand muon propagation to the experimental site at 4850 ft below the surface we performed detailed simulations with Geant4 \cite{Allison2006,Agostinelli2003250}, version 4.96p04, using the QGSP\_BIC\_HP physics list with muon-nuclear processes turned on. A surface map of the area, 10 km in radius, surrounding  the laboratory was implemented in \GF\ with a granularity of 77 x 100 m, see Fig.~\ref{fig:landscape}. This large simulated area allows entry angles between 0 and 78 degrees relative to the vertical axis for muons entering the underground laboratory. Muons with energies between 5 GeV and 500 TeV using parametrization from  \cite{Guan} were generated randomly on a 100 km$^2$ plane at an altitude of 2500 m using the surface muon flux energy and angular distribution from Ref. \cite{Hansen2003}, and their propagation through the rock was recorded. 

For the detector response component of the simulations we used both the GEANT3 package \cite{G3} in addition to Geant4 to check for consistency. We used energy and angular distributions of muons entering our laboratory from the muon propagation simulations in order to determine the effective area ($A_{\rm eff}$) for muons detected in both coincidence configurations. We generated 271,000  muons over an area of $10\times10$ m$^2$, which is much larger than that of the veto array. This surface at which the muon paths were initiated was situated 1~m above the rock ceiling of the laboratory, 2 meters above the upper panels. These muons were then propagated through the laboratory, and events in which more than 1 MeV is deposited in a panel by either the muon or its secondaries were recorded. All details of the \DEM\ shielding were included in the simulation model. 

For the two-fold-coincidence configuration simulation, 8779 muons were recorded, resulting in an effective area ($A_{\rm eff}$) of $3.24\times10^4$ cm$^2$. For the three-fold-coincidence configuration simulation, 2876 muons were detected resulting in $A_{\rm eff}=1.15\times10^4$ cm$^2$.

The muon flux ($F$) (Eq. \ref{eq-flux}) is calculated using $A_{\rm eff}$,  the number of muons observed ($N_{obs}$), and the live time $T$ of each configuration. The statistical uncertainties are large enough that the systematic uncertainties are negligible. 
\begin{equation}
F = \frac{N_{obs}}{A_{\rm eff} \epsilon T}
\label{eq-flux}
\end{equation}
The coincidence detectrion efficiency $\epsilon$ is taken to be $>$99.7\% based on the single-panel efficiency ($\epsilon_D$) measured in~\cite{Bugg2014}. For the first configuration with two-fold coincidence, the reconstructed flux was found to be $(5.09\pm0.24)\times10^{-9}$~$\mu$/s/cm$^2$. For the second configuration with three-fold coincidence, the reconstructed muon flux was found to be $(5.54\pm0.23)\times10^{-9}$~$\mu$/s/cm$^2$. Although there were fewer muons registered for the second configuration, the statistical accuracy is similar to the first configuration due to the absence of the random coincidence background from $\gamma$ rays. It should also be noted that data were taken with the first configuration when the muon flux was near its annual minimum, while the second configuration data were taken near the annual maximum flux. The 4-5 percent level annual variation of the muon flux is on the same order as our present statistical sensitivity, and will be the subject of future study.

Combining results from both measurements gives a total muon flux of $(5.31\pm0.17)\times10^{-9}$~$\mu$/s/cm$^2$, taken to be an average over the seasonal variation. These two results derive from two very different geometries and angular acceptances. With agreement better than one sigma, we conclude that the statistical uncertainty dominates. 

\begin{figure}[ht]
\begin{center}
\includegraphics[width=9cm]{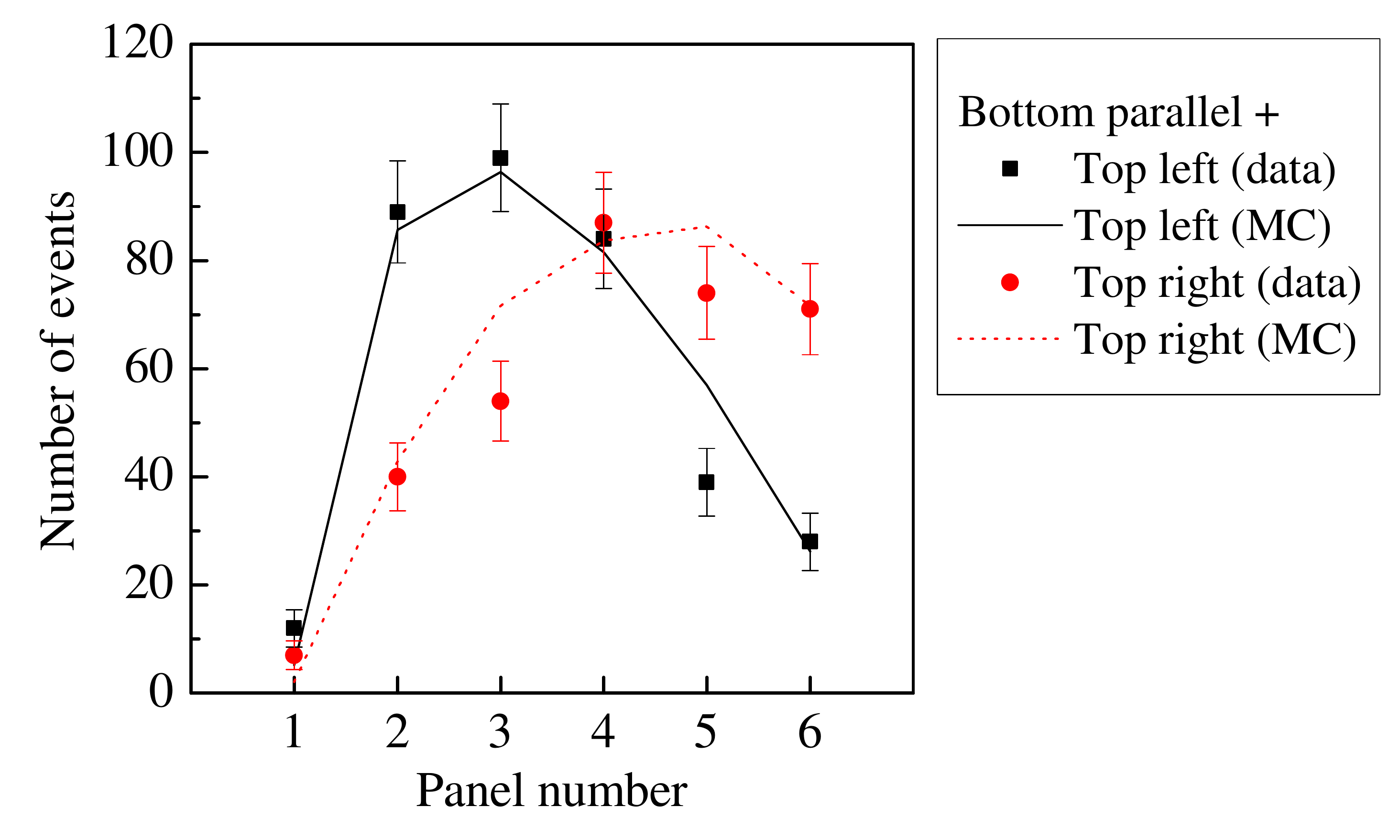}
\caption{Color online: Event rate for the coincidence between the bottom 6 panels and the upper 2 panels for the three-fold coincidence configuration. The simulation predictions are shown as lines and data are displayed with statistical error bars.}
\label{fig:AngDist}
\end{center}
\end{figure}

The installed detector configuration did not permit a study of the muon angular dependence with high angular resolution. Nevertheless, it was possible to compare the angular distribution between data and simulation for the three-fold coincidence configuration by using the hit pattern in the bottom narrow panels relative to the coincident top panel. In Fig.~\ref{fig:AngDist}, the event rate for the coincidence between the bottom six panels and two top panels is shown. The top left veto panel is located over bottom panels 7, 8, and 9, and the top right veto panel is situated over bottom panels 10, 11, and 12. For panels 7-12, the numbering indicates sequential position from left to right. The bottom panel array is shifted $\sim$20 cm to the right relative to that of the two top panels. The distance between the top and bottom planes is about 2 meters. There is good agreement between simulation and data within the existing statistical precision.

\section{Discussion}

Our measured total flux is somewhat larger than the calculation in Ref.~\cite{Mei2006}, although the two agree at the 2-$\sigma$ level. Reference~\cite{Mei2006} approximated the SURF overburden with a flat surface profile. 

An early measurement~\cite{Cherry1983} of the vertical muon flux resulted in a value of $(4.91 \pm 0.06)\times10^{-9}$~$\mu$/s/cm$^2$/sr. This measurement employed large water Cherenkov tanks ($200\times200\times120$ cm$^3$) stacked in 3 layers. Events consisting of coincident signals within 3 tanks in a vertical-path trajectory corresponding to an effective zenith angle $<$18 degrees were selected for analysis. To compare our estimate for the total muon flux and that of Ref.~\cite{Cherry1983}, we integrate our total flux within an 18-degree cone. We calculate a vertical muon flux of $(4.42\pm0.15_\textrm{stat.})\times10^{-9}$~$\mu$/s/cm$^2$/sr using our own muon model where the stated uncertainty is only the statistical uncertainty from our total flux measurement. However, the vertical flux extracted from other muon models based on our measured total flux predict different values of $(4.16\pm0.12_\textrm{stat.})\times10^{-9}$~$\mu$/s/cm$^2$/sr  using Ref.~\cite{Zhang2015} and $(5.05\pm0.16_\textrm{stat.})\times10^{-9}$~$\mu$/s/cm$^2$/sr  using a muon angular distribution~\cite{Vitaly} derived from the MUSIC package~\cite{MUSIC}. The spread in the extracted vertical fluxes is a result of differences in the angular distributions near small zenith angle and is indicative of a systematic uncertainty in the overburden model and in the simulations. Taking the standard deviation of the three as an estimate of the systematic uncertainty, we calculate the vertical flux to be $(4.4\pm0.7_\textrm{syst.})\times10^{-9}$~$\mu$/s/cm$^2$/sr. The total muon flux, on the other hand, is insensitive to the choice of angular distribution model -- the systematic uncertainty in the total flux extracted using the three different angular distributions is negligible relative to the statistical uncertainty.

We would like to note two things, however, in comparing our extracted vertical flux relative to the vertical flux measurement of Ref.~\cite{Cherry1983}. First, the quoted uncertainties in Ref.~\cite{Cherry1983} are entirely statistical and no systematic uncertainty was estimated. We were unable to obtain from the authors additional details about the Ref.~\cite{Cherry1983} geometry and thresholds in order to simulate their apparatus with the different muon flux angular profiles. Second, while the measurement of \cite{Cherry1983} was performed at the same underground level, the separate location of the two experiments beneath the sharp surface profile results in slightly different overburdens and azimuthal muon flux distributions. Nonetheless, when including the systematic uncertainty in the muon models due to differing rock density and angular distribution, our calculated vertical muon flux is consistent with Ref. \cite{Cherry1983}. 

\begin{figure}[ht]
\begin{center}
\includegraphics[width=9cm]{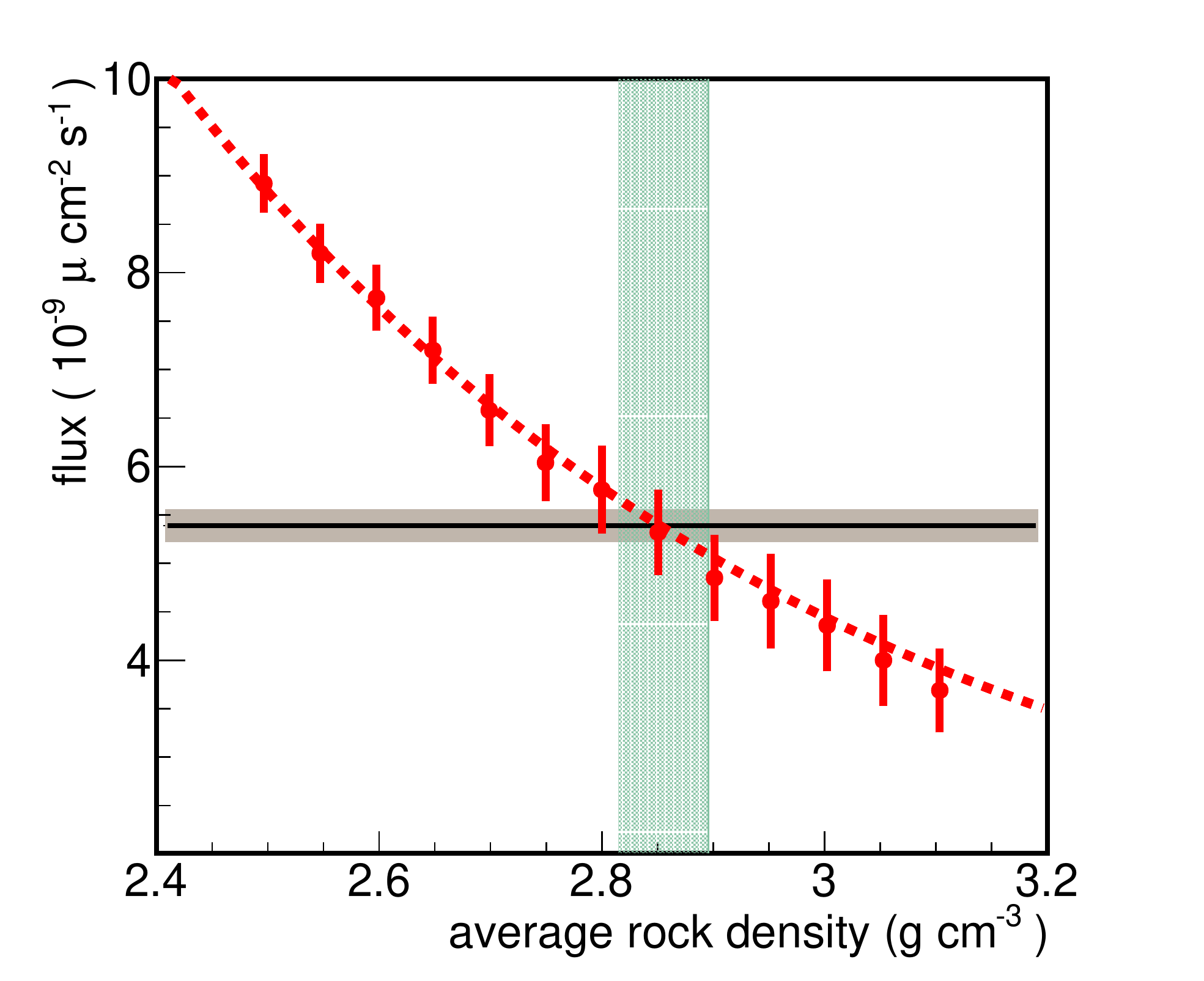}
\caption{Color online: The predicted muon flux at 4850 ft based on the Geant4 simulation described in the text for several values of the average rock density. The dashed (red) curve is an exponential fit to those simulated data points. The total simulation uncertainties are indicated by the error bars and are correlated between neighboring points. The grey shaded region represents our measurement confidence interval with the central value indicated by the black line. The vertical green band shows rock density range from geological studies.}
\label{fig:density}
\end{center}
\end{figure}

To study the effect of rock density further, the total simulated muon flux at the Davis Campus was evaluated over a range of rock densities. The simulated flux has to be normalized to the muon flux at the surface, so experimental data is used to derive a scaling factor. A surface muon flux~\cite{Hansen2003} of $2.0~\pm~0.2$~$\mu$/s/cm$^2$ is used as reference. The uncertainties in this value take into account the uncertainty in altitude as well as possible seasonal variations of the atmospheric temperature resulting in a variation of the muon flux~\cite{Dmitrieva2011401,Borexino2013}.
The dependance of rock density on the total flux can be seen in Fig.~\ref{fig:density}. We find that a rock density of 2.89 $\pm$ 0.06  g/cm$^3$ yields a total muon flux consistent with our measurement. This result agrees very well with geological studies at SURF that found an average rock density of 2.86 $\pm$ 0.11 g/cm$^3$~\cite{Heise2015} (taking the nominal 4\% uncertainty) based on cone 45 degrees from vertical.

\section{Conclusion}
We report for the first time a measurement of the total flux of muons at the SURF Davis Campus. This flux is necessary for present and future experiments to assess cosmic-ray induced backgrounds at this underground location. A measured total flux permits such an assessment with less interpretation than would be required to incorporate effects of the rock density, surface topology, and muon angular distribution. Previous measurements were done at the 800 and 2000 ft levels~\cite{Gray2011}. The measured flux was found to be in good agreement with that predicted in~\cite{Mei2006} and with our own simulations using a rock density similar to values measured in geological studies. A comparison of our result with an older measurement of the vertical flux ~\cite{Cherry1983} is consistent when including a systematic uncertainty on the muon angular distribution needed to convert our total flux into a vertical flux. The \MJ\ \DEM\ veto system is operating in the underground environment and identifies muons as expected.

\section*{Acknowledgments}
We thank the technical contributions of Tom Burritt, Greg Harper, Rick Huffstetler, Randy Hughes, Eric Olivas, Alvin Peak II, David Peterson, Larry Rodriguez, Harry Salazar, Jared Thompson, and Tim Van Wechel. We would also like to thank Vitaly Kudryavtsev for useful discussions.

This material is based upon work supported by the U.S. Department of Energy, Office of Science, Office of Nuclear Physics under Award  Numbers DE-AC02-05CH11231, DE-AC52-06NA25396, DE-FG02-97ER41041, \\ DE-FG02-97ER41033, DE-FG02-97ER41042, DE-SC0012612, \\ DE-FG02-10ER41715,  DE-SC0010254, and DE-FG02-97ER41020. We acknowledge support from the Particle Astrophysics Program and Nuclear Physics Program of the National Science Foundation through grant numbers PHY-0919270, PHY-1003940, 0855314, PHY-1202950, MRI 0923142 and 1003399. We acknowledge support from the Russian Foundation for Basic Research, grant No. 15-02-02919. We  acknowledge the support of the U.S. Department of Energy through the LANL/LDRD Program.  We thank our hosts and colleagues at the Sanford Underground Research Facility for their support.

\section{References}
\bibliographystyle{iopart-num.bst}
\bibliography{MuonFlux.bbl}

\end{document}